\documentstyle[aps, preprint]{revtex}
\tightenlines

\def\A{\mbox{\bf A}}
\def\x{\mbox{\bf x}}

\def\Tr{\mbox{\rm Tr}}
\def\n{n}
\def\be{\begin{equation}}
\def\ee{\end{equation}}

\begin{document}
\input epsf
\draft
\preprint{PURD-TH-00-06, quant-ph/0010072}
\date{October 2000; revised December 2000}

\title{Decoherence in a superconducting ring}
\author{S. Khlebnikov}
\address{
Department of Physics, Purdue University, West Lafayette, IN 47907, USA
}
\maketitle

\begin{abstract}
A superconducting ring has different sectors of states corresponding to 
different values of the trapped magnetic flux; this multitude of states
can be used for quantum information storage. If a current supporting
a nonzero flux is set up in the ring, fluctuations of electromagnetic 
field will be able to ``detect'' that current and thus cause a loss
of quantum coherence. We estimate the decoherence exponent for a ring of
a round type-II wire and find that it contains a macroscopic suppression 
factor $(\delta / R_{1})^{2}$, where $R_{1}$ is the radius of the wire, 
and $\delta$ is the London penetration depth. We present some encouraging
numerical estimates based on this result.
\end{abstract}

\pacs{PACS numbers: 03.67.Lx}
\section{Introduction and results}
A superconducting ring is a good candidate for macroscopic quantum memory 
\cite{qm}. Its ability to store quantum information stems from the 
complicated ``vacuum structure'', viz.
the presence of different sectors of states corresponding to different 
values of magnetic flux through the ring.

It is easy to suppress spontaneous transitions between these
different sectors by making the ring sufficiently thick \cite{qm}. In the
absence of such transitions, 
the main mechanism by which quantum information stored in the ring 
deteriorates
is {\em decoherence}, i.e. a decrease, in time, of matrix elements
between states with different values of the flux. Decoherence is due to
interaction of the macroscopic variable, in our case the supercurrent in 
the ring, 
with the ``environment'' comprised by local degrees of 
freedom---fluctuations of 
the electromagnetic field. 

The supercurrent density required to support a single flux quantum through
a macroscopic ring is very small. 
This makes that state {\em locally} almost indistinguishable 
from the state with no flux at all.
Based on this observation, we have suggested
\cite{qm} that in this case decoherence may not be as strong as it is 
generally believed to be for macroscopic systems.
The purpose of the present paper is to discuss decoherence in 
the ring quantitatively.

In general, the environment
starts out in an arbitrary initial state with some density matrix $\rho_0$.
Two different values of the macroscopic variable give rise to two
different histories of the environment, which are represented by 
evolution operators $U_1(t)$ and $U_2(t)$. We define the amount of
coherence (at time $t$) between these two histories as the ``overlap''
\be
{\cal C}(t) = \Tr[ U_1(t) \rho_0  U_2^{\dagger}(t)] \; .
\label{olap}
\ee
Next, we define a decoherence exponent $D(t)$ by
\be
 \exp[-D(t)]  = \left| {\cal C}(t) \right|^2 \; .
\label{Dgen}
\ee

For an environment comprised by linear oscillators, which in addition 
interact linearly with the macroscopic variable, decoherence had been widely 
discussed in connection with macroscopic tunneling \cite{CL,ChL}. 
That problem is similar but not identical to ours: in quantum
memory tunneling is suppressed. Indeed, our problem is more analogous to the
``error-free'' case of ref. \cite{Unruh}; however, we will be able to
supply definite values for all of our parameters.

In principle, the linear approximation can break down at large times
even for weakly nonlinear environments. In addition, whatever we designate
as the environment will interact with a still larger system and will loose
energy to that system at some characteristic dissipation rate. For example,
in our case, the electromagnetic field can be absorbed on the surface
of the superconductor, as well as on the surfaces of any surrounding 
objects. In such cases, the evolution of the environment can no
longer be represented by unitary operators.

It is certainly of interest to understand how
nonlinearity and dissipation in the environment affect quantum coherence
in the original system. However, for our present purposes 
the linear approximation will be sufficient. 
A mechanism for decoherence exists already at the linear level \cite{ChL}. 
Dissipation will only be discussed phenomenologically,
to make sure that it does not significantly alter the evolution 
of the oscillators at relevant times.

Specifically, we consider the following problem.
Suppose at the initial time $t=0$ a supercurrent has been set up in the ring
but the electromagnetic field is still in the thermal state corresponding 
to zero current, with a Gibbs distribution $w_m$ of probabilities for the energy
eigenstates $|m\rangle$. Because the field is displaced from its true
equilibrium (corresponding to a nonzero average magnetic field),
it will oscillate. Consider two oscillation histories, corresponding
to two different values of the current (one of which may be zero), and call 
their evolution operators $U_1(t)$ and $U_2(t)$. The decoherence exponent 
in this case will be determined by
\be
 \exp[-D(t)] = \sum_m \left|
\langle m| U_2^{\dagger} U_1 |m\rangle \right|^2 w_m \; .
\label{D}
\ee
We describe a calculation of $D(t)$ below.

The problem just posed is admittedly somewhat artificial. Imagine that 
we switch from
the state with no flux (and no current) to the state with unit flux by 
dragging
a single flux line through the bulk of the superconductor. There is no 
reason to
expect that the current will equilibrate first, and the field 
next, as in
our model problem. Nevertheless, we think that it is good to know the 
answer to 
this model problem before turning to more realistic setups. 

The answer is this. Consider a circular ring of radius $R_0$ made out of 
a London-type superconducting round wire of radius $R_1$, in the limit
\be
\delta \ll R_1 \ll R_0 \; ,
\label{limit}
\ee
where $\delta$ is the London penetration depth. 
We find that at intermediate times $R_1/c \ll t \ll R_0/c$ ($c$ is the speed
of light), the exponent
$D(t)$ grows as $t$ for $t \ll \hbar / k_B T$ and as $t^2$ for
$t \gg \hbar / k_B T$, up to some logarithms. At large times, $t \gg R_0/c$, 
the exponent saturates at a certain
limiting value, for which we obtain the following estimate:
\be
D_{\rm lim} \sim \frac{f(u)} {16\alpha_{\rm EM}\ln^4(R_0/R_1)}
\left( \frac{\pi c}{R_0 \omega_{\rm min}} \right)
\left(\frac{\delta}{R_1}\right)^2 \; ,
\label{Dlim}
\ee
where $\alpha_{\rm EM}$ is the fine-structure constant, and
$u = k_B T /\hbar\omega_{\rm min}$ with $\omega_{\rm min} \sim c/R_0$; 
$f(u) = 1$ for small $u$, and $f(u) = u$ for large $u$.

The main result is the macroscopic suppression factor 
$(\delta / R_{1})^{2}$ in (\ref{Dlim}).
Taking for estimates
$\delta \sim 10^{-5}$ cm, $R_{0}= 1$ cm, $R_{1} = 1$ mm, 
$\omega_{\rm min} = \pi c / R_0 \sim 10^{11}$ s$^{-1}$, and replacing the
logarithm (which controls logarithmic accuracy) with unity, we obtain
$x\sim T/({\rm 1~K})$ and $D_{\rm lim} \sim 10^{-7} f(x)$.

These results apply not only to a ring made out of a solid 
superconductor, but also to a ring coated with a superconducting film,
as long as the thickness of the film is much larger than the London
depth $\delta$. Similar calculations can be done for any type
of persistent circular current, of which a supercurrent in a macroscopic
ring is but one limiting case. The opposite limit is
a circular current in a single atom---e.g. a circular Rydberg state
\cite{circular}, which has already been proposed as a basis for quantum
computation \cite{Domokos&al}. It would be interesting to do the calculation
for that case and see how the decoherence rate changes as the current loop
becomes microscopic.

We now outline the main steps leading to estimate (\ref{Dlim}).

\section{The oscillator Hamiltonian}
We will only consider decoherence at times
\be
t \gg R_{1} / c \; ,
\label{times}
\ee
when, as we will see, the main contribution to the decoherence
exponent comes from electromagnetic modes with low frequencies:
\be
\omega \ll c / R_1 \; .
\label{omega}
\ee
For such modes, the response of a superconductor of the London type
(the only type we consider) is determined, to a good accuracy,
by the London penetration depth $\delta$.

To count the oscillator modes, we enclose the ring in a large
normalization sphere of radius ${\cal R} \gg R_0$, centered at the center
of the ring. We define a system of spherical coordinates
$(r,\theta,\phi)$, with angle $\theta$ measured form the axis of the
ring's rotational symmetry.

Only the azimuthal component of the vector potential, indeed only
the $\phi$-independent modes of it, interact with the current.
For these modes, the vector potential takes the form
\be
\A(r,\theta,\phi; t) = \mbox{\bf e}_{\phi}(\phi) A(r,\theta; t) \; ,
\label{vec}
\ee
where $\mbox{\bf e}_{\phi}$ is the azimuthal unit vector. Then,
\be
\nabla^2 \A = \mbox{\bf e}_{\phi} \left(
\nabla^2 - \frac{1}{r^2\sin^2\theta} \right) A \; ,
\label{nabla2}
\ee
and the wave equation for ${\bf A}$ takes the form
\be
\ddot{A} - \nabla^2 A +  \left[
\frac{1}{r^2\sin^2\theta} + V(r,\theta) \right] A 
= \frac{4\pi}{c} j \; .
\label{eqm}
\ee
Here $j$ is the constant (in time) supercurrent density, and 
$V(r,\theta)$ is the ``potential'' that represents the Meissner effect: 
inside the superconductor
\be
V(r,\theta) = 1 / \delta^2 \; ,
\label{Vx}
\ee
while outside the superconductor both $j$ and $V$ are zero.

The equation of motion (\ref{eqm}) can be obtained, after substitution
(\ref{vec}), from the effective Hamiltonian
\be
H = \int d^3 x \left( \frac{1}{8\pi c^2} \dot{\A}^2 
+ \frac{1}{8 \pi} \mbox{\bf B}^2 
+\frac{V}{8\pi} {\A}^2 - \frac{1}{c} \A \mbox{\bf j} \right) \; ,
\label{H}
\ee
where $\mbox{\bf B}=\nabla\times\A$ is the magnetic field, and 
$\mbox{\bf j} = \mbox{\bf e}_{\phi} j$.
Such Hamiltonian formulation neglects any dissipation (absorption)
of the electromagnetic field. This approximation is justified in
Appendix A. We stress again that decoherence 
we calculate in this paper is not due to any such dissipation, but
is a result of a ``measurement'' of the supercurrent, performed by
fluctuations of the electromagnetic field.

Now, expand the field $A$ in a complete set of modes $f_\n(r,\theta)$:
\be
A(r,\theta; t) = \sum_{\n} f_{\n}(r,\theta) X_{\n}(t) \; .
\label{X}
\ee
This expansion defines a set of oscillator coordinates $X_\n$. The modes 
$f_\n$
satisfy the following equation:
\be
\left[ -\nabla^2 + \frac{1}{r^2\sin^2\theta} + V(r,\theta) \right] 
f_{\n} = 
\frac{\omega_{\n}^2}{c^2} f_{\n} \; ,
\label{modeq}
\ee
where $\omega_n$ are real eigenfrequencies, together with the
continuity conditions at the surface of the superconductor and
some boundary condition at the boundary of the normalization volume. They
are normalized in that volume by
\be
\int d^3 x f_{\n}(r,\theta) f_{\n'} (r,\theta) = \delta_{\n\n'} \; .
\label{norm}
\ee
A typical low-frequency mode is shown schematically
in Fig. 1. Notice that its main support is outside the superconductor.

Substituting the expansion (\ref{X}) in the Hamiltonian (\ref{H}) we 
obtain the
Hamiltonian for the oscillators:
\be
H = \frac{1}{8\pi c^2} \sum_{\n} \left( \dot{X}_{\n}^2 + \omega_{\n}^2 
X_{\n}^2
\right) 
- {1\over c} \sum_{\n} X_{\n} \int d^3 x j(r,\theta) f_{\n}(r,\theta) \; ,
\label{H1}
\ee
which is of the Caldeira-Leggett \cite{CL} type. 
The role of a macroscopic variable is played by
the total supercurrent $I$. Because in our case the total
supercurrent has no dynamics of its own,
it is not necessary to split each integral in (\ref{H1}) into a product 
of $q\equiv I$ and a coupling $C_{\n}$. Nevertheless, to conform to the 
established 
notation, we will denote these integrals as
\be
q C_{\n} = -\frac{1}{c} \int d^3 x j(r,\theta) f_{\n}(r,\theta) \; .
\label{C}
\ee

The quantum-mechanical problem specified by the Hamiltonian (\ref{H1}) is 
exactly solvable, and the overlap (\ref{olap}) is easily calculable.
For the decoherence exponent of two states corresponding to two 
different values $q=q_1$ and $q=q_2$, with the initial state being a thermal 
ensemble at $q=0$, we obtain
\be
D(t) = \frac{2 (\Delta q)^2}{\pi \hbar} \int_0^\infty d\omega
\frac{J(\omega)}{\omega^2} [ 1 - \cos\omega t] \coth(\hbar\beta\omega /2) 
\; ,
\label{D1}
\ee
where $\Delta q= q_1 - q_2$, $\beta = 1/ k_B T$, and $J(\omega)$ is 
the spectral density defined as in refs. \cite{CL,ChL}:
\be
J(\omega) = \frac{\pi}{2} \sum_{\n} \frac{C_{\n}^2}{m_{\n} 
\omega_{\n}}
\delta(\omega_{\n} - \omega) \; ;
\label{J}
\ee
according to (\ref{H1}) the masses of all oscillators are
\be
m_{\n} = \frac{1}{4\pi c^2} \; .
\label{m}
\ee
Eq. (\ref{D1}) is similar to expressions found in the literature 
\cite{ChL},
but we should stress that ours is {\em not} a macroscopic tunneling 
problem: 
as we pointed 
out before, in quantum memory, we strive to prevent spontaneous 
transitions
(tunneling or otherwise) between states with different values of $q$. 
These
different values now simply label various macroscopically distinct 
histories.

Expression (\ref{D1}) applies at both zero and finite temperatures. 
Two limiting cases can be considered: $k_B T \ll \hbar \omega$, and
$k_B T \gg \hbar \omega$, where $\omega$ is a typical frequency
at which the integral in (\ref{D1}) saturates. 
In the first case (``low'' temperatures),
\be
D(t) = \frac{2 (\Delta q)^2}{\pi \hbar} \int_0^\infty d\omega
\frac{J(\omega)}{\omega^2} [ 1 - \cos\omega t] \; ;
\label{lowT}
\ee
in the second case (``high'' temperatures),
\be
D(t) = \frac{4 (\Delta q)^2}{\pi \hbar^2 \beta} \int_0^\infty d\omega
\frac{J(\omega)}{\omega^3} [ 1 - \cos\omega t] \; .
\label{D2}
\ee
To complete the calculation of $D$ we now need to find the couplings
$C_\n$, for the low-frequency modes of interest.

\section{Decoherence at intermediate times}
Our strategy will be as follows. We first compute the spectral density
$J(\omega)$ for $\omega$ in the range
\be
c/R_0 \ll \omega \ll c/R_1 \; .
\label{range}
\ee
The form of the spectral density at these $\omega$ determines the decoherence 
exponent at intermediate times
\be
R_1/c \ll t \ll R_{0}/c \; .
\label{times1}
\ee
Then, in the next section, we establish that at smaller 
$\omega$, $\omega \ll c/R_0$, $J(\omega)$
is proportional to $\omega^3$. According to (\ref{D1}), that means that
at $t\gg R_0/c$ the exponent $D(t)$ does not grow beyond a certain
limiting value $D_{\rm lim}$. 
Finally, we estimate $D_{\rm lim}$.

For each mode of oscillation, we define a wavenumber $k_n$ by
\be
k_n = \omega_n / c \; .
\label{kn}
\ee
At any time in the range (\ref{times1}), the main contribution to 
the decoherence exponent comes from modes with
\be
k_n \sim 1 / c t \; .
\label{low}
\ee
For counting these modes, the normalization volume can be replaced
by a tube, coaxial with the
wire, of radius $R_{\rm norm}$ that satisfies
\be
ct \ll R_{\rm norm} \ll R_{0} \; .
\label{Rt}
\ee
This new normalization volume is more convenient than the original
sphere because the problem now acquires an 
approximate cylindrical symmetry: we can view the wire as a straight
cylinder, with fields subject to periodic boundary conditions at the ends.

This cylindrical problem has a natural set of cylindrical coordinates, 
$(\rho, \varphi, z)$. The expression (\ref{vec}) for the vector potential
can now be approximated by
\be
\A(\rho,\varphi;t) = \mbox{\bf e}_z A(\rho,\varphi;t) \; ,
\label{vec1}
\ee
and the mode equation (\ref{modeq}) by
\be
\left[ -\nabla^2 + V(\rho) \right] f_{\n}(\rho,\varphi) = 
\frac{\omega_{\n}^2}{c^2} f_{\n}(\rho,\varphi) \; .
\label{modeq1}
\ee
As seen from (\ref{modeq1}), the mode functions $f_n$ are eigenfunctions of
$-i\partial/\partial\varphi$, the angular momentum about the $z$ axis.
We concentrate on functions with zero angular momentum (i.e. those independent
of the angle $\varphi$), because for other values the coefficients
$C_n$ in (\ref{J}) are suppressed by additional powers of $k_n R_1$. Thus, 
we are left with $f_n$ that depend only on the radial
coordinate $\rho$, the distance to the wire's axis.

By assumption, the wire's radius $R_1$ is much larger than the 
penetration depth 
$\delta$. Then, the solution to equation (\ref{modeq}) inside
the superconductor is approximately
\be
f_{\n}(\rho) \approx A_{\n} \exp[(\rho - R_1) / \delta] \; .
\label{in1}
\ee
On the outside, the solution is a combination of Bessel functions; near 
the
boundary, for $k_n$ satisfying (\ref{low}), it reduces to
\be
f_{\n}(\rho) \approx A_{\n} + B_{\n} \ln (\rho / R_1) \; .
\label{out1}
\ee
Eq. (\ref{out1}) already takes into account the continuity of the 
solution on the
surface; the continuity of the first derivative requires 
$B_{\n} = A_{\n} R_1 / \delta$, so (\ref{out1}) becomes
\be
f_{\n}(\rho) \approx A_{\n} \left( 1 + \frac{R_1}{\delta} \ln (\rho / 
R_1) 
\right) \; .
\label{out2}
\ee
Notice the enhancement of the coefficient of the logarithm; it comes from 
matching the first derivatives.

The coefficient $A_{\n}$ in (\ref{in1}), (\ref{out2}) is determined from the 
normalization 
condition (\ref{norm}).
To compute the normalization integral, we need the solution further away 
from the
surface of the superconductor, where the form (\ref{out2}) no longer 
applies. 
But (\ref{out2}) does allow us to pick up the coefficients of the Bessel 
functions.
Writing $\ln (\rho / R_1) = \ln (k_n \rho) - \ln (k_n R_1)$, we observe 
that the
coefficient of $Y_0(k_n \rho)$ (the function of the second kind) is 
proportional to
$A_n R_1 / \delta$, while the coefficient of $J_0(k_n \rho)$ (the function 
of the first
kind) is additionally enhanced by $\ln (k_n R_1)$. 

We will be content with logarithmic
accuracy; to such accuracy, away from the surface we have
\be
f_{\n}(\rho) =  - A_{\n} \frac{R_1}{\delta}  \ln (k_n R_1) J_0(k_n \rho) 
\; .
\label{out3}
\ee
To count the modes, we use the Dirichlet boundary condition at
$\rho = R_{\rm norm}$:
\be
f_{\n}(R_{\rm norm}) = 0 \; .
\label{bou}
\ee
Normalizing (\ref{out3}) in a cylinder of radius $R_{\rm norm}$ and 
length $L$, 
we find
\be
A_n = \left( \frac{k_n}{2 L R_{\rm norm}} \right)^{1/2} \frac{\delta}{R_1}
\frac{1}{|\ln k_n R_1|} \; .
\label{An}
\ee

Finally, using for $j$ the equilibrium current density 
\be
j(\rho) = \frac{I}{2\pi R_1 \delta} \exp[(\rho - R_1) / \delta] \; ,
\label{j}
\ee
where $I$ is the total supercurrent, we obtain for the spectral density
\be
q^2 J(\omega) = \frac{\pi}{4 c^2} \left( \frac{\delta}{R_1} \right)^2
\frac{I^2 L}{\ln^2 (\omega R_1 / c)} \; .
\label{Jfin}
\ee

We now specialize to the case when $q_1 = 0$, and
$q_2=I_s$ is the current corresponding to one quantum of flux through the 
ring: 
to logarithmic accuracy
\be
I_s = \frac{c \Phi_0}{2L\ln(L/R_1)} \; ,
\label{Is}
\ee
where $\Phi_0 = \pi \hbar c / e$ is the flux quantum. We obtain
\be
(\Delta q)^2 J(\omega) = \frac{\pi^3 c^2 \hbar^2}{16 e^2 L \ln^2(L/R_1)}
\left( \frac{\delta}{R_1} \right)^2
\frac{1}{\ln^2(\omega R_1 / c)}
\label{JI}
\ee
($e$ is the absolute value of the electron charge).

With this form of $J(\omega)$, the integral in eq. (\ref{D1}) gets the main
contribution from $\omega \sim 1 /t$. So, at
$t \ll \hbar / k_B T$, we apply eq. (\ref{lowT}), to obtain
(to logarithmic accuracy)
\be
D(t) = \frac{t \pi^3 c^2 \hbar}{16 e^2 L \ln^2(L/R_1)}
\left( \frac{\delta}{R_1} \right)^2
\frac{1}{\ln^2(ct / R_1)} \; .
\label{lowT1}
\ee

At $t \gg \hbar / k_B T$, we apply the ``high''-temperature limit
(\ref{D2}).
Although the integral in (\ref{D2}) in our case is convergent, we impose
a low-frequency cutoff at $\omega = c / R_0$,
a representative value below which eq. (\ref{Jfin}) is not applicable.
This will give us an estimate of the accuracy of the result. We obtain
\be
D(t) = \frac{t^2 \pi^2 c^2 k_B T}{8 e^2 L\ln^2(L/R_1)} 
\left( \frac{\delta}{R_1} \right)^2
\left[ \frac{1}{\ln(ct/R_1)} - \frac{1}{\ln(R_0/R_1)} 
+ O \left( \frac{1}{\ln^2(ct/R_1)} \right) \right] \; .
\label{Dfin}
\ee
The second term in the bracket is a correction due to the low-frequency
cutoff.
Under the condition (\ref{times1}), this term is small compared to 
the first term (the one from $\omega \sim 1/t$), but as
$t$ approaches values of order $R_0 / c$ it becomes comparable to the first
term. Thus, at $t \sim R_0/c$ the expansion indicated in 
(\ref{Dfin}) breaks down, and we need a different method to establish 
the form of $D(t)$ at large $t$.

\section{Decoherence at large times}
We now turn to modes with 
\be
k_{n} \ll 1/R_{0} \; . 
\label{lowest}
\ee
For counting these, we use the original normalization sphere and
the Dirichlet boundary condition
\be
f_n({\cal R}) = 0 \; .
\label{bou2}
\ee
The modes are normalized in the sphere by the condition (\ref{norm}).

Let us expand each mode function $f_n$ in associated Legendre functions
$P_l^1(\cos\theta)$:
\be
f_n(r, \theta) = (2\pi N_l)^{-1/2} \sum_{l\geq 1} S_{nl}(r) 
P_l^1(\cos\theta) \; ,
\label{expP}
\ee
where $S_{nl}$ are some radial functions, and $N_l$ is the normalization
integral for $P_l^1$:
\be
N_l = \frac{2l(l+1)}{2l + 1} \; .
\label{Nl}
\ee
Note the absence of $l=0$: there is no (nonzero) $P_0^1$.

Eq. (\ref{norm}) now translates into the
following normalization condition for the radial functions $S_{nl}$:
\be
 \sum_{l\geq 1} \int_0^{{\cal R}} S_{nl}(r) S_{n'l}(r) r^2 dr 
= \delta_{nn'} \; ,
\label{normS}
\ee
while the mode equation (\ref{modeq}) leads to the following
equation for $S_{nl}$:
\be
-\frac{1}{r^2} \frac{d}{dr} \left( r^2 \frac{d}{dr} S_{nl} \right)
+\frac{l(l+1)}{r^2} S_{nl} - k_n^2 S_{nl} + \sum_{l'} V_{ll'} S_{nl'} = 0 
\; .
\label{eqS}
\ee
The ``potential'' $V_{ll'}$ here is given by
\be
V_{ll'}(r) = (N_l N_{l'})^{-1/2} 
\int_{0}^{\pi} V(r, \theta) P_l^1(\cos\theta) P_{l'}^1(\cos\theta)
\sin\theta d\theta \; ,
\label{Vll'}
\ee
in terms of the ``potential'' $V(r, \theta)$.

The potential (\ref{Vll'}) is nonzero only for $R_{\rm in} < r < R_{\rm 
out}$,
where $R_{\rm in}$ and $R_{\rm out}$ are the inner and outer radii of the 
ring.
In the region $r < R_{\rm in}$, 
eq. (\ref{eqS}) is a free radial equation, in which we can now neglect 
$k_n^2$.
A solution regular at $r=0$ is then
\be
S_{nl}(r) = D_{nl} r^l \; ,
\label{reg}
\ee
where $D_{nl}$ are some constants. 
In the outside region, $r > R_{\rm out}$, where eq. (\ref{eqS}) is also 
free, the solution is a linear combination of two spherical waves. As 
long as $r \ll 1/k_{n}$, so that $k_n^2$ is still negligible, we can 
write
\be
S_{nl}(r) = F_{nl} \left( \frac{r}{R_{\rm out}} \right)^{l}
+ G_{nl} \left( \frac{R_{\rm out}}{r} \right)^{l+1} \; ,
\label{lc}
\ee
with some constants $F_{nl}$ and $G_{nl}$. On the other hand, at large
distances $r \gg 1/k_n$, we have (cf. ref. \cite{LL})
\be
S_{nl}(r) \approx 
\frac{F_{nl} (2l+1)!!}{r k_n^{l+1} R_{\rm out}^l \cos\delta_n}
\sin(k_n r - \pi l/2 + \delta_n) \; ,
\label{shift}
\ee
where $\delta_n$ is a phase shift.

Because the smallest possible $l$ is $l=1$, from the normalization
condition (\ref{normS}) we see that
\be
F_{nl} = O(k_n^2)
\label{Fnl}
\ee
(the phase shift $\delta_n$ is $O(k_n^3)$).
For each $n$, the sets of coefficients $F_{nl}$, $G_{nl}$, and
$D_{nl}$ are related through scattering on the potential $V_{ll'}$.
Because this scattering occurs at $r\sim R_0 \ll 1/k_n$, it is
insensitive to $k_n$, so $G_{nl}$ and $D_{nl}$, and therefore
also $f_n$ on the surface of the wire, are all $O(k_n^2)$.
Then, according to eq. (\ref{J}), the spectral density $J(\omega)$ is
$O(\omega^3)$.

This suppression of the spectral density at small $\omega$ can be
explained by noting that the large wavelength fluctuations of the field
``see'' simultaneously two diametrically opposite segments of the wire,
with currents that add up to zero. Such fluctuations therefore have only
derivative interactions with the current; hence the extra powers of
$k_n$, or $\omega$.

From eq. (\ref{D1}), we now see that the low frequency region
$\omega \ll c / R_0$ does not significantly contribute to decoherence,
and the main contribution at $t \gg R_0/c$ comes from modes with
$\omega \sim c / R_0$. We can estimate decoherence at these large times
by using (\ref{D1}) with the expression (\ref{JI}), which correctly
describes higher frequencies, and an infrared cutoff at
$\omega_{\rm min} \sim c / R_0$. In this way, we arrive at the estimate
(\ref{Dlim}).

\appendix
\section{The role of dissipation}
Here we list some estimates for absorption rates of
low-frequency electromagnetic field on the surface of a superconductor. 
The purpose is to show that the corresponding dissipation time is
much larger than the timescale $R_0/c$, at which the decoherence
exponent saturates.

The response kernel $Q$, which determines 
the current induced in the superconductor by the electromagnetic field,
is defined by
\be
j_{\rm ind}(\x, t) = -\int d^3 \x' dt' Q(\x, \x'; t- t') A(\x', t') \; .
\label{jind}
\ee
The low-frequency expansion of the Fourier transform of $Q$ 
(inside the superconductor) is
\be
Q(\omega) = \frac{1}{4\pi\delta^2} - i\omega \sigma / c^2 + \ldots \; ,
\label{Q2}
\ee
where the first term is the one taken into account in eq. (\ref{H}),
while the second term describes dissipation due to a finite 
conductivity $\sigma$. 

We only consider low frequencies,
\be
\omega \ll c / R_1 \; ,
\label{omax}
\ee
and conventional ($s$-wave) superconductors with critical temperatures 
$T_c$ of order of a few Kelvin. For $R_1 \sim 1$ mm, we have 
$\hbar\omega \ll 2 \Delta_0$, where $\Delta_0$ is the zero-temperature gap.
So, at sufficiently low temperatures, a single photon cannot break a Cooper 
pair, and the dissipation is entirely due to thermally
excited quasiparticles. Therefore, we estimate the conductivity as
\be
\sigma \sim \sigma_N \exp[-\Delta(T) / k_B T] \; ,
\label{sigma}
\ee
where $\sigma_N$ is the normal-state conductivity at $T$ around $T_c$,
and $\Delta(T)$ is the temperature-dependent gap.

Using (\ref{sigma}) with $\sigma_N \sim 10^{18}~{\rm s}^{-1}$ 
(corresponding to $10^8 ~\Omega^{-1} {\rm m}^{-1}$ in SI) 
and $\delta \sim 10^{-5}$ cm, 
we find that for our range of frequencies 
the second term in (\ref{Q2}) is much smaller than the first. 
This allows us to expand the surface impedance as
\be
\zeta(\omega) = \omega [-4\pi c^2 Q(\omega)]^{-1/2} = 
-i \frac{\omega\delta}{c} \left[ 1 + 2\pi i \omega \sigma \delta^2 / c^2 +
O(\omega^2) \right] \; .
\label{zeta}
\ee
The dissipative effect (absorption) is represented by the real part 
of $\zeta$,
\be
\zeta_R(\omega) = 2\pi \omega^2 \sigma \delta^3 / c^3 
\sim  10^{-6} (\omega / 10^{11}~{\rm s}^{-1})^2 \exp[-\Delta(T) / k_B T] \; ,
\label{zetaR}
\ee
where the estimate is
for the same values of $\sigma_N$ and $\delta$ as before.

For example, suppose that the device is inside a cavity, and the cavity 
and the ring are made of the same superconducting material. Then,
assuming that radiation does not leak out,
dissipation is mainly due to absorption on the walls of the cavity.
The rate of dissipation is estimated as
\be
\tau^{-1} \sim \zeta_R c / R_{\rm cav} \; ,
\label{tau1}
\ee
where $R_{\rm cav} > R_0$ is the radius of the cavity. The timescale $\tau$ 
is thus $1/\zeta_R$ times larger than $R_{\rm cav} /c$, which in turn is
larger than $R_0/c$, the timescale at which decoherence saturates.

\begin{figure}
\leavevmode\epsfxsize=5in \epsfbox{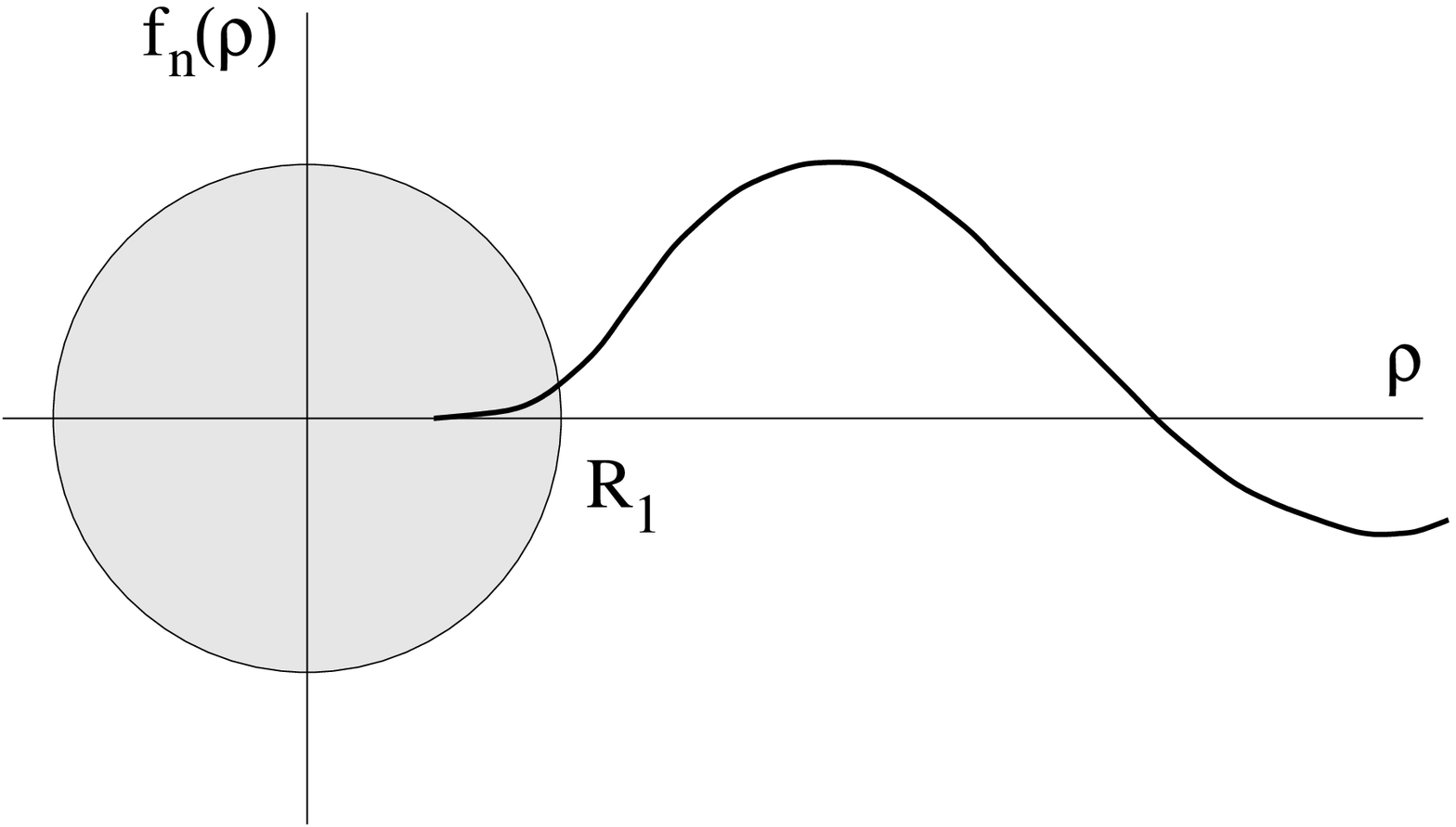}
\vspace*{0.2in}
\caption{A sketch of the mode function $f_n$ corresponding to a low-frequency
mode of the electromagnetic field, in the presence of a round superconducting 
wire (of radius $R_1$). The sketch is meant to reflect two features:
the main support of $f_n$ is outside 
the superconductor, and the value of $f_n$ on the surface is much smaller than
its typical value outside.
}
\label{fig:mode_function}
\end{figure}
\end{document}